# Molecular states observed in a single pair of strongly coupled self-assembled InAs quantum dots

T. Ota[1], M. Stopa[1], M. Rontani[2], T. Hatano[1], K. Yamada[1], S. Tarucha[1,3,4],
H. Z. Song[5], Y. Nakata[5], T. Miyazawa[5], T. Ohshima[5] and N. Yokoyama[5]

[1]*Mesoscopic Correlation Project, ERATO, JST, Atsugi-shi, Kanagawa 243-0198, Japan*
[2]*INFM-S3 and Dipartimento di Fisica, Universita degli Studi di Modena e Reggio Emilia, Modena, Italy*
[3]*University of Tokyo, Bunkyo-ku, Tokyo, 113-0033, Japan*
[4]*NTT Basic Research Laboratories, Atsugi-shi, Kanagawa 243-0198, Japan*
[5]*Fujitsu Laboratories Ltd., Atsugi-shi, Kanagawa 243-0197, Japan*

**Abstract**

Molecular states in a *single pair* of strongly coupled self-assembled InAs quantum dots are investigated using a sub-micron sized single electron transistor containing just a few pairs of coupled InAs dots embedded in a GaAs matrix. We observe a series of well-formed Coulomb diamonds with charging energy of less than 5 meV, which are much smaller than those reported previously. This is because electrons are occupied in molecular states, which are spread over both dots and occupy a large volume. In the measurement of ground and excited state single electron transport spectra with magnetic field, we find that the electrons are sequentially trapped in symmetric and anti-symmetric states. This result is well-explained by numerical calculation using an exact diagonalization method.

*Corresponding author. Present address: Tarucha Mesoscopic Correlation Project, NTT Atsugi Research and Development Center, 4S-308S, 3-1, Wakamiya, Morinosato, Atsugi-shi, 243-0198, Japan
*E-mail address:* ta-ota@tarucha.jst.go.jp
Tel: +81- 46-248-4000, Fax: +81-46-248-4014

Semiconductor coupled quantum dots are often referred to as artificial molecules since their electronic properties resemble those of real molecules, in which both tunnel coupling and Coulomb interaction between the coupled dots are important parameters. The artificial molecules have a "covalent" or "ionic" type bonding, depending on strength of the quantum mechanical coupling between the dots. The competition between these two mechanisms has been investigated in lithographically defined coupled dots [1] and coupled self-assembled InAs quantum dots [2].

The coupled self-assembled InAs quantum dots are of great interest for viewpoint of fundamental physics as well as applications for opto-electronic devices such as laser, detector, quantum computing and so on. The molecular states in the single pair of them have been intensively studied by photoluminescence spectroscopy [3]. In contrast, there are only a few works on electron transport through the single pair of them, though this technique can directly provide information of zero-dimensionality, interaction effects and interdot quantum mechanical coupling for electrons in this system.

In previous reports, we have succeeded to fabricate a sub-micron sized single-electron transistor containing a few pairs of the coupled InAs dots and distinguish the single-electron transport through a *single* pair of them by exactly controlling number of electrons in the dots, starting from *zero*. Using this technique, we have found marked differences in the transport properties between a single pair of the strongly and weakly coupled dots [2,4]. In this work, using strongly coupled dots, we investigate single-electron tunneling through the molecular states.

Figure 1(a) schematically shows the cross-section of the sample, i.e. single-electron transistor, prepared for the experiment. The material consists of a 750 nm-thick n-doped GaAs buffer layer on GaAs (100) semi-insulating substrate, a 33 nm i-GaAs layer incorporating two InAs self–assembled quantum dot layers in the center and a 400-nm thick n-doped GaAs layer. There is a 2 nm thick AlGaAs layer on both ends of the i-GaAs layer. The InAs self-assembled dots are formed on an InAs wetting layer, and the two wetting layers are 11.5 nm apart. The typical height, and diameter of the dots are 2 nm, and 20 nm, respectively. The average density of the dots is on the order of $10^9$ cm$^{-2}$. The prepared samples have a cylindrical pillar with geometrical diameter of 0.35 µm, surrounded by Schottky gate metal. Detail of the sample structure is shown in Ref. [2]. About a few coupled dots are present inside the pillar. Transport measurements on these samples are performed at 20 mK in dilution refrigerator.

Although about two or three coupled dots exist in the pillar, we are usually able to distinguish the transport through a single pair of them in the non-linear transport regime using *position-sensitive* single electron spectroscopy technique. When two or more pairs of the coupled dots are located close together, different families of Coulomb diamonds, bounded by threshold lines with different slopes, are observed in the non-linear transport regime [2]. This indicates that gate capacitance to each pair of the coupled dots strongly depends on their location [5].

Figure 1(b) shows a gray log-scale plot of *dI/dV* as a function of source-drain voltage ($V_{SD}$) and gate voltage ($V_G$). A series of well-formed Coulomb diamonds with charging energy of several meV are observed along $V_{SD}$=0 V. The electron number N varies from N=10 at $V_G$= -1.55 V to N=0 at the pinch off voltage of $V_G$= -2.5 V. The estimated charging energy is much smaller than those estimated by capacitance-voltage characteristics in the InAs dot ensembles [6]. This result indicates that the electrons are occupied in the coherent molecular states, which are spread over both dots and occupy a large volume.

The evolution of the current of the single-electron transport spectra, i.e. Coulomb oscillation, with magnetic field provides information of the orbitals in which the electrons are trapped. Figure 2(a) shows a log-scale plot of Coulomb oscillation as a function of magnetic field with $V_{SD}$=0.5 meV. The peaks are seen to evolve in pairs, implying that each single-particle state is filled with two electrons of opposite spin. In order to analyze these current peaks, we consider Fock-Darwin (F-D) states $E_{n,l}$, assuming that the dots are confined by the parabolic potential. Here, *n* and *l* are the radial quantum number and angular momentum quantum number, respectively. Figure 2(b) shows calculation of F-D states of the *single dot* with quantum confinement energy $\hbar\omega_0$ of 10 meV [7].

Each single-particle state can occupy two electrons with opposite spin and they form spin-pairs. By comparing Figs. 2(a) and (b), the current peaks of N=1 to N=4 and those of N=5 and N=6 exhibit $E_{0,0}$ and $E_{0,1}$ like behavior, respectively. This result can't be explained using F-D spectra of the single dot and indicate that the peaks of N=1 to N=4 originate from the electron filling in the molecular states, i.e. symmetric and anti-symmetric states, of the s-orbital. In the current peaks $N \geq 7$ in Fig. 2(a), we can see several kinks at ~ 2 T, which comes from crossing between single particle states.

Now, we focus on the current peaks of N=1 to N=4 of the s-orbital and perform excitation spectroscopy of them in order to study further the excited states as well as the ground states. Figure 3(a) shows a gray log-scale plot of $dI/dV_G$ as a function of the magnetic field with $V_{SD}$= 7 meV. For N=1, we find an excited state, which runs parallel to the ground state. For N=2, an excited state crosses to the ground state at B= ~ 12 T. For N=3 and N=4, the evolutions of the ground and excited states with magnetic field are more complicated. Several transitions between the excited states and ground states are recognized, shown as the triangles in Fig. 3(a). In order to investigate what kind of states are responsible for the ground and excited states in Fig. 3(a), we perform numerical calculation. Figure 3(b) shows numerical calculation using an exact diagonalization method [8]. The upper and lower panels correspond to the peaks of N=3-4 and N=1-2, respectively. The ground states correspond to the bold lines. By comparing Figs. 3(a) and (b), we find that the ground and excited states in Fig. 3(a) come from the electron injection in the symmetric and anti-symmetric states of s-orbital. The energy spacing between these two peaks directly reflects to the energy splitting between symmetric and anti-symmetric states, $\Delta_{SAS}$. About 4 meV of $\Delta_{SAS}$ is extracted from Fig. 3(a). The excited state of N=2 peaks can be assigned to the triplet state of the s-orbital. The excited states in the experimental data is not so clear and it is not easy to assign the spin configurations. Considering the ground states of N=1 to N=4 below B=~ 4 T, we have found the electrons are subsequently occupied first in the symmetric state and then in the anti-symmetric state.

In conclusion, we have demonstrated single-electron charging spectroscopy in a single pair of strongly coupled InAs dots with precise manipulation of the energy levels of the dots. We have found the electrons are accumulated in the symmetric and anti-symmetric states, which is analyzed by the numerical calculation using the exact diagonalization method.

The authors thank K. Ono, S. Amaha and T. Sato for valuable discussions. The authors acknowledge financial supports from the Grant-in-Aid for Scientific Research A (No. 40302799), CREST-JST, Focused Research and Development Project for the Realization of the World's Most Advanced IT Nation, IT program, MEXT and MIUR-FIRB "Quantum phases...", INFM I.T. Calcolo Parallelo 2003, and the Italian Minister for Foreign Affairs, DGPCC.

[Figure captions]

Figure 1

Schematic illustration of cross-section of the single-electron transistor including a few coupled InAs dots (a) and Gray log-scale plot of $dI/dV_{SD}$ as a function of source-drain voltage and gate voltage, showing the transport through a single pair of the coupled dots at 20 mK (b).

Figure 2

Gray log-scale plot of $dI/dV_G$ of the Coulomb oscillations with $V_{SD}$=0.5 mV as a function of gate voltage and magnetic field at 20 mK (a) and calculation of Fock-Darwin spectra of the *single dot* assuming 10 meV of the quantum confinement energy (b).

Figure 3

Gray log-scale plot of $dI/dV_G$ of the Coulomb oscillations with $V_{SD}$=7 mV as a function of gate voltage and magnetic field at 20 mK (a) and numerical calculation using an exact diagonalization (b) The upper and lower panels correspond to the peaks of N=3-4 and N=1-2, respectively.

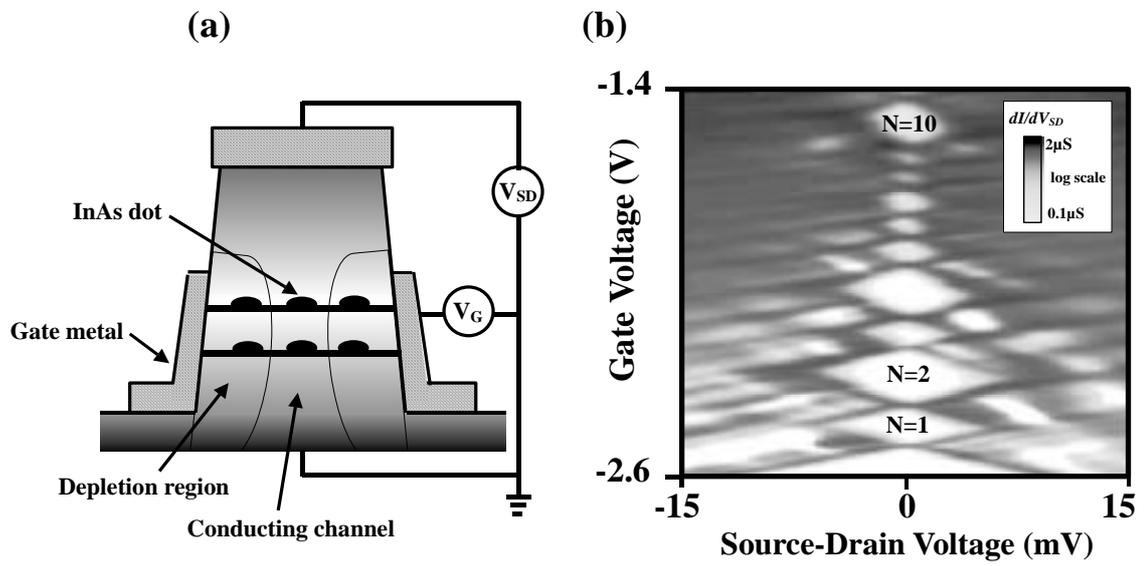

**Figure 1**

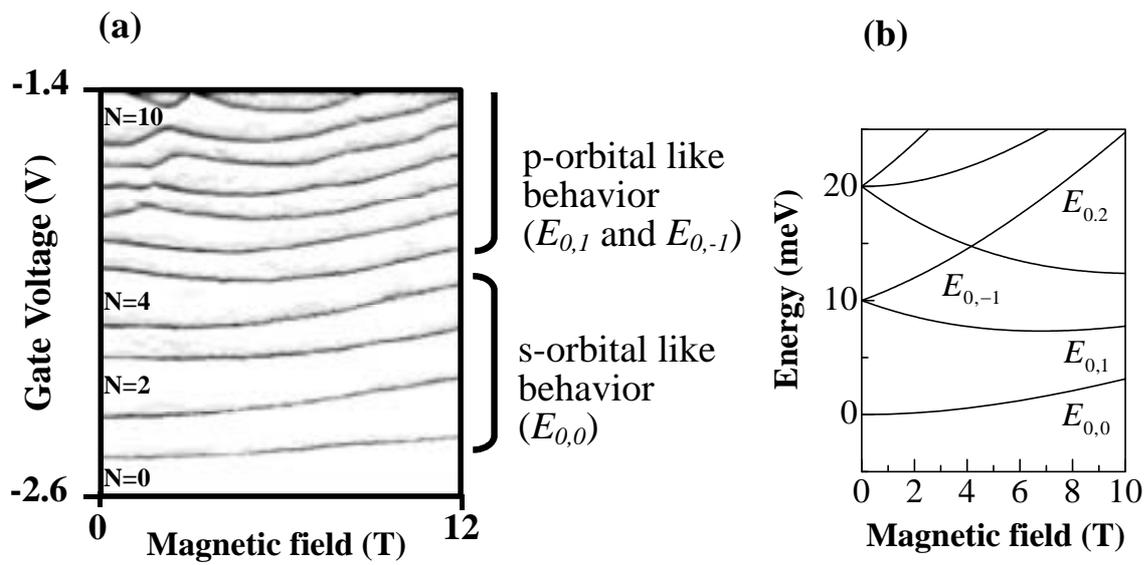

**Figure 2**

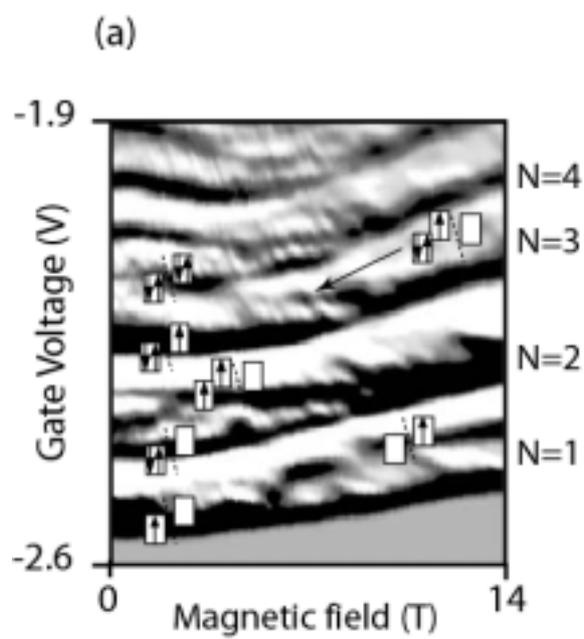 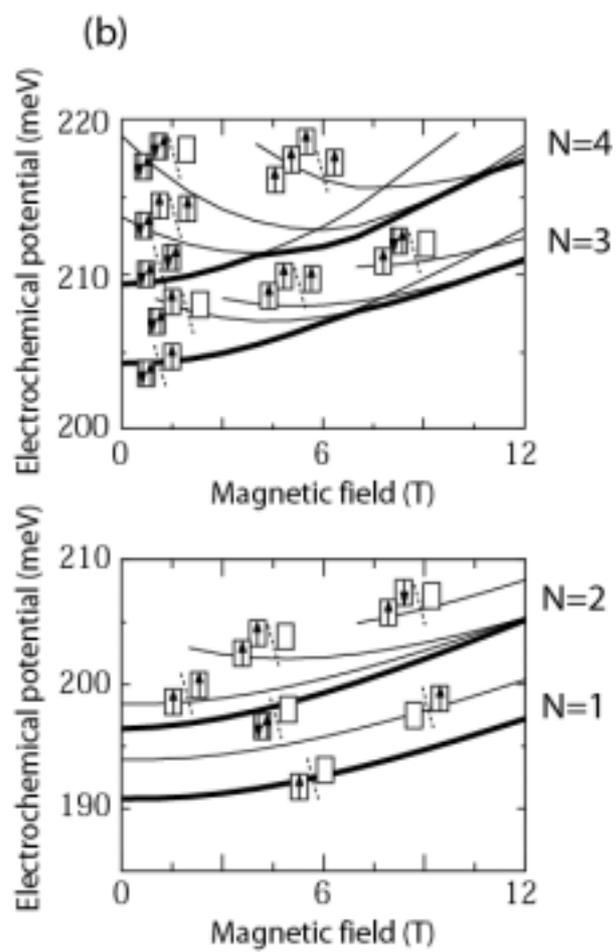